\documentclass[aps,prl,twocolumn,a4paper,showpacs,superscriptaddress,floatfix,10pt]{revtex4-1}

\usepackage{amsthm,amsmath,amsfonts,amssymb,verbatim, color}
\usepackage{graphicx}
\usepackage{subfigure}
\usepackage{float}
\usepackage{epsfig,slashed}
\usepackage{helvet}
\usepackage{epstopdf}
\usepackage{bbold}
\usepackage{cancel}
\usepackage{color}
\usepackage[colorlinks=true,citecolor=blue,
linkcolor=blue,urlcolor=blue]{hyperref}
\usepackage[nodisplayskipstretch]{setspace}

\newcommand{\ket}[1]{\ensuremath{\left| #1 \right\rangle}}
\newcommand{\bra}[1]{\ensuremath{\left\langle #1 \right|}}

\newcommand{\averag}[1]{\ensuremath{\langle #1 \rangle}}

\newcommand{\abs}[1]{\ensuremath{\vert #1 \vert}}

\newcommand{\Ryd}{\ensuremath{\mathcal{R}}}

\def\up{\uparrow}
\def\down{\downarrow}

\begin{document}

\title{Floquet Symmetry-Protected Topological Phases in Cold-Atom Systems}

\author{I.-D. Potirniche}
\affiliation{Department of Physics, University of California, Berkeley, California, 94720, USA}

\author{A. C. Potter}
\affiliation{Department of Physics, The University of Texas at Austin, Austin, Texas 78712, USA}

\author{M. Schleier-Smith}
\affiliation{Department of Physics, Stanford University, Stanford, California 94305, USA}

\author{A. Vishwanath}
\affiliation{Department of Physics, University of California, Berkeley, California, 94720, USA}
\affiliation{Department of Physics, Harvard University, Cambridge, MA 02138, USA}

\author{N. Y. Yao}
\affiliation{Department of Physics, University of California, Berkeley, California, 94720, USA}

\date\today

\begin{abstract}

We propose and analyze two distinct routes toward realizing interacting symmetry-protected topological (SPT) phases via periodic driving. First, we demonstrate that a driven transverse-field Ising model can be used to engineer complex interactions which enable the emulation of an equilibrium SPT phase. This phase remains stable only within a parametric time scale controlled by the driving frequency, beyond which its topological features break down. To overcome this issue, we consider an alternate route based upon realizing an intrinsically Floquet SPT phase that does not have any equilibrium analog. In both cases, we show that disorder, leading to many-body localization, prevents runaway heating and enables the observation of coherent quantum dynamics at high energy densities. Furthermore, we clarify the distinction between the equilibrium and Floquet SPT phases by identifying a unique micromotion-based entanglement spectrum signature of the latter. Finally, we propose a unifying implementation in a one-dimensional chain of Rydberg-dressed atoms and show that protected edge modes are observable on realistic experimental time scales.

\end{abstract}

\maketitle

The discovery of topological insulators---materials which are insulating in their interior but can conduct on their surface---has led to a multitude of advances at the interface of condensed matter physics and materials engineering~\cite{bernevig2006quantum,chen2009experimental,hasan2010colloquium,qi2011topological,bernevig}. At their core, such insulators are characterized by the existence of nontrivial topology in their underlying single-particle electronic band structure~\cite{kane2005z,moore2007topological}. Generalizing our understanding of topological phases to the presence of strong many-body interactions represents one of the central questions in modern physics. Some of the simplest generalizations that have emerged along this direction are symmetry-protected topological (SPT) phases~\cite{senthil_spt, ashvin_spt_review, chen_classification_2011}, which represent the minimal extension of topological band insulators to include many-body correlations. 
Featuring short-range entanglement, SPT phases do not exhibit anyonic excitations in their bulk, but nevertheless possess protected edge modes on their surface; as a result, they represent a particularly fertile ground for studying the interplay between symmetry, topology, and interactions. 

While indirect signatures of certain ground state SPTs have been observed in the solid state~\cite{buyers86,morra88,xu96}, directly probing the quantum coherence of their underlying edge modes represents an outstanding experimental challenge. In principle, cold-atom quantum simulations could offer a powerful additional tool set---including locally-resolved measurements and interferometric protocols---for probing the robustness of edge modes and systematically exploring their stability to specific perturbations~\cite{bakr2009quantum,atala2013direct,cheuk2015quantum,haller2015single,parsons2016site}. Moreover, such platforms could also enable the controlled storage and transmission of quantum information~\cite{sondhi_mbl_order,bahri_localization_2015,yao_many-body_2015}. Despite these advantages, and owing to the complexity of typical model SPT Hamiltonians, it remains difficult to engineer and stabilize SPT phases in cold-atom systems. 

\begin{figure}
\centering
\includegraphics[width=1.\columnwidth]{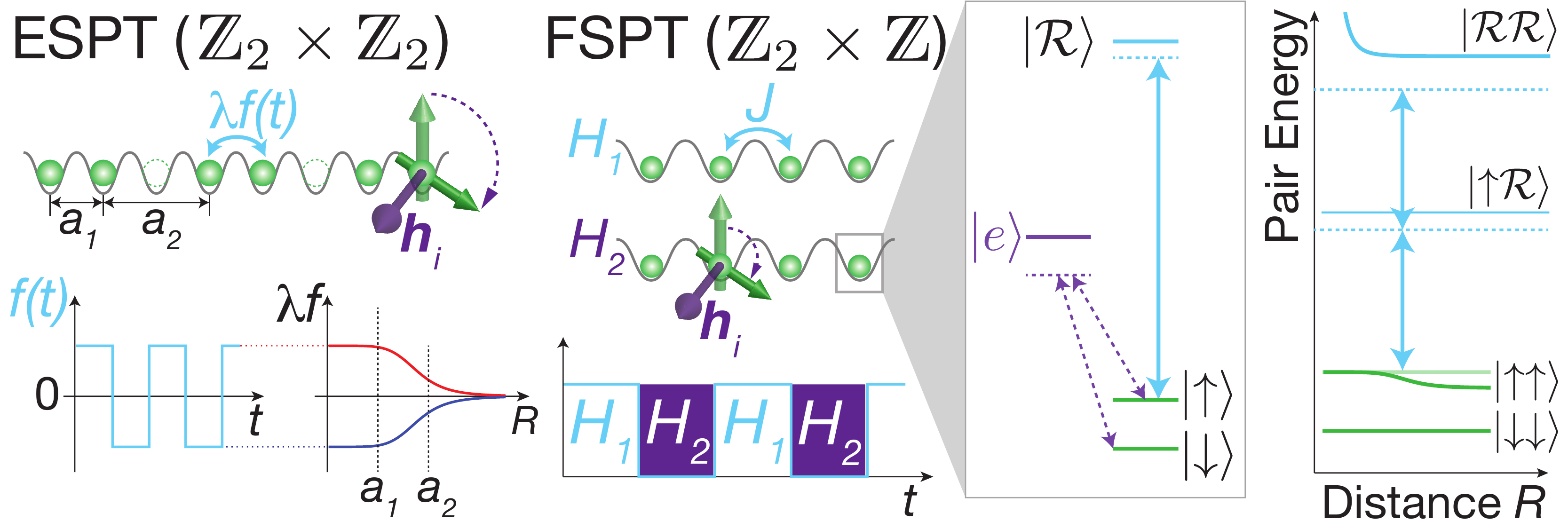}
\caption{A 1D array of atoms is trapped in an optical lattice or tweezer array. Ising interactions for pseudo-spin states $\ket{\down},\ket{\up}$ are generated by optically coupling $\ket{\up}$ to Rydberg state $\ket{\Ryd}$ (solid blue arrows).  Random fields $h_i$ are generated by a spatially varying Raman coupling (dotted purple arrows) between $\ket{\down}$ and $\ket{\up}$.  While emulating the ESPT phase requires a dimerized chain with Ising couplings $\lambda f(t)$ of dynamically switchable sign, the FSPT phase is simulated simply by alternating between two Hamiltonians consisting of Ising interactions ($H_1$) and a disordered transverse field ($H_2$).}
\label{fig:schematic}
\end{figure}

One approach to this challenge is to emulate the complex interactions giving rise to static, equilibrium SPT (ESPT) phases by periodically driving a simpler Hamiltonian at frequencies much larger than its intrinsic energy scales~\cite{Iadecola_stroboscopic_2015}. In addition to this approach, seminal results on classifying driven (Floquet) phases~\cite{khemani_phase_2015,sondhi_floquet_class1,sondhi_floquet_class2, vishwanath_floquet_class, nayak_floquet_class, harper_floquet_class} have also shown that there exist Floquet-SPT (FSPT) phases which are inherently dynamical and have no static analog. Interestingly, such a FSPT phase can be realized at driving frequencies that are \emph{comparable} to the energy scales of the bare Hamiltonian.

The power of periodic driving for engineering topological phases has been extensively explored  in cold-atom~\cite{aidelsburger11,jotzu14,kennedy15}, solid-state~\cite{oka09,refael11,gedik13}, and photonic~\cite{rechtsman13,titum15}  systems.  For cold atoms, where Floquet control has so far been applied only to single-particle band structures~\cite{Cooper11,aidelsburger11,hauke12,jotzu14,kennedy15,flaschner16}, recent advances in optically controlling interactions~\cite{Clark15,Gaul16,Jau16,Zeiher16,Goldschmidt16,Labuhn16,zeiher17,lukin17} offer new opportunities for accessing strongly correlated phases~\cite{yao2013realizing,vanBijnen15,Glaetzle15, barbiero16}.  Notably, coherent spin-spin interactions with a range of several microns~\cite{Jau16,Zeiher16,zeiher17,lukin17} can be introduced via Rydberg dressing~\cite{Rolston10,Henkel10,Pupillo10,Gil14,Jau16,Zeiher16,Goldschmidt16,zeiher17,lukin17}.  However, prospects for modulating such dressing light in order to Floquet engineer many-body Hamiltonians has remained largely unexplored.

This owes, in part, to the difficulty of generating quantum coherent order in an interacting Floquet system which will typically absorb energy from the driving field, eventually heating to a featureless infinite temperature state~\cite{rigol14,lazarides_equilibrium_2014}. This difficulty is further exacerbated for isolated  atomic systems, where the lack of coupling to an external bath renders the system incapable of releasing excess energy and entropy~\cite{stamper2009}. A fruitful strategy for combating such heating is to harness many-body localization (MBL)~\cite{huse_localization-protected_2013, nandkishore_many-body_2015,altman_many-body_2015, ponte_floquet_2015,khemani_phase_2015}, which has been predicted to stabilize quantum coherent behavior without the need for stringent cooling or adiabatic preparation of low temperature many-body states~\cite{bahri_localization_2015, sondhi_mbl_order, yao_many-body_2015,barkeshli2015continuous}. 

In this Letter, we propose to exploit periodically driven interactions to realize two distinct non-equilibrium MBL SPT phases in a one-dimensional array of cold atoms  (Fig.~\ref{fig:schematic}).  Driving the interaction term of a transverse-field Ising model (TFIM) enables the emulation of an ESPT phase whose edge modes are protected by an emergent $\mathbb{Z}_2\times\mathbb{Z}_2$ symmetry~\cite{Iadecola_stroboscopic_2015}. This phase remains stable only within a parametric time scale controlled by the driving frequency, beyond which its topological features break down. Alternatively, toggling between Hamiltonians with solely Ising interactions or purely transverse fields yields an intrinsically dynamical FSPT phase which has no equilibrium analog.  We explore the stability of both phases to long-range interactions and provide a detailed  experimental blueprint using  Rydberg-dressed atoms.

\emph{ESPT phase}---Inspired by pioneering work on emulating static phases in driven systems~\cite{oka09,kitagawa_topological_2010,refael11,jiang_majorana_2011,Iadecola_stroboscopic_2015,torre16,rudner_anomalous_2013,tonylee16}, we first consider the realization of a many-body localized version of the Haldane phase~\cite{haldane_nonlinear_1983}. This SPT phase can be protected by a discrete dihedral symmetry, $\mathbb{Z}_{2}\times\mathbb{Z}_{2}$, and exhibits boundary modes that are odd under the symmetry; these edge modes behave as decoupled spin-1/2 degrees of freedom that are robust to any perturbation which preserves the symmetry.

We begin by examining the robustness of the edge modes in a periodically driven and dimerized spin chain (Fig.~\ref{fig:schematic}):
\begin{equation}
\label{eq:drivenHamiltonian}
H_{0}(t) = \sum_{i=1}^{N} h_{i} \sigma_{i}^{x} + \sum_{i=1}^{N-1}  f(t) \lambda_{i}\sigma_{i}^{z}\sigma_{i+1}^{z} + V_{x}\sigma_{i}^{x}\sigma_{i+1}^{x},
\end{equation}
where $N$ represents an even number of spins, $\sigma_{i}^{\alpha}$ are the Pauli operators on site $i$, $\lambda_{2k+1} = \lambda_{1}$,  $\lambda_{2k} = \lambda_{2}$  (with $\lambda_{1}, \lambda_{2}>0$) and  $f(t) = \omega \cos(\omega t)$ is the driving function \footnote{Any driving function that changes sign every half-period will suffice. For instance, tweaking the binary drive from~\cite{khemani_phase_2015} to have this property would also yield an ESPT.}.  For $V_{x} = 0$, the model  is noninteracting and exhibits edge dynamics which never decohere~\cite{Iadecola_stroboscopic_2015}.  Here, we first verify that the SPT phase remains stable under the addition of short-range interactions $V_x\neq 0$ that preserve the  dihedral symmetry (generated by products of $\sigma^{x}_i$ on the even and odd sites).  We then assess the effects of more generic, longer range, interactions.

\begin{figure}[t]
\centering
\includegraphics[width=\columnwidth]{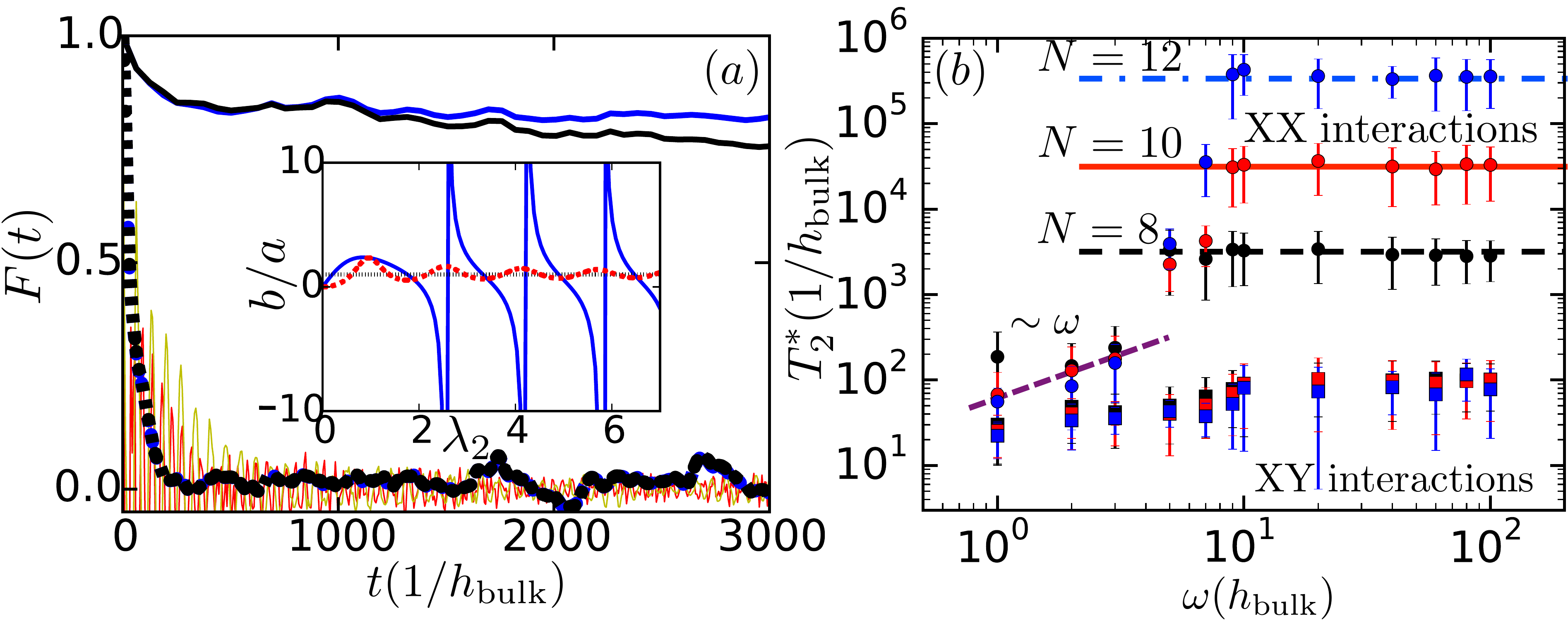}
\caption{ESPT phase---(a) $F^{\alpha}(t)$ for $N=10$ spins with $\omega = 100$, $V_{x} = 0.05$, $V_{y} = 0$,  $\lambda_{1} = 1.54$ and $\lambda_{2} =0.69$, yielding $b(\lambda_{1}, \lambda_{2})/a(\lambda_{1}, \lambda_{2}) \sim 10$.
Almost overlapping dotted lines represent the clean undisordered case (black and blue for $F^{z}$ and $F^{x}$, respectively). Solid lines correspond to strong on-site disorder, with thick black and blue lines for $F^{z}$ and $F^{x}$ in the dimerized case and thin solid yellow and red lines for $F^{z}$ and $F^{x}$ in the undimerized case.
(inset) Ratio $b(1, \lambda_{2})/a(1, \lambda_{2})$ in the dimerized (solid blue) and the undimerized (dotted red) models. The SPT phase corresponds to $b/a > 1$ (delimited by the dotted black line).
(b) $T_{2}^{*}$ as a function of frequency and system size~\cite{suppinfo}. As $\omega$ is increased for $V_{x}=0.05$ (circles), $T_{2}^{*}$ saturates consistent with being bounded by  $T_{2}^{*} \sim \min(\mathcal{O}(\omega), e^{\mathcal{O}(N)})$. Adding generic interactions, $V_{y}\sum_{i}\sigma_{i}^{y}\sigma_{i+1}^{y}$ with $V_{y}=0.2$ (squares), leads to a breakdown of the edge coherence for all parameters.}
\label{fig:ESPT}
\end{figure}

In the limit of large driving frequencies $\omega$, the dynamics are described by an effective time-independent Floquet Hamiltonian, $H_{\mathrm{F}}$, which can be constructed perturbatively in orders of $1/\omega$ using a Magnus expansion~\cite{abanin_floquet_2015_v1,kuwahara16,abaninprethermal}. At leading order, we obtain the time-averaged Floquet Hamiltonian~\cite{suppinfo}
\begin{eqnarray}\label{eq:stroboscopicHamiltonian}
H^{(0)}_{\mathrm{F}} &=& \sum_{i=1}^{N} h_{i}a(\lambda_{1}, \lambda_{2}) \sigma_{i}^{x} - \sum_{i=2}^{N-1}h_{i}b(\lambda_{1}, \lambda_{2}) \sigma_{i-1}^{z}\sigma_{i}^{x}\sigma_{i+1}^{z}\\ \nonumber
&+& V_{x}J_{0}(2\lambda_{2})(\sigma_{1}^{x}\sigma_{2}^{x} + \sigma_{N-1}^{x}\sigma_{N}^{x})\\ \nonumber
&+& V_{x}\sum_{i=2}^{N-2}\left[c(\lambda_{i+1})\sigma_{i}^{x}\sigma_{i+1}^{x} +d(\lambda_{i+1})\sigma_{i-1}^{z}\sigma_{i}^{y}\sigma_{i+1}^{y}\sigma_{i+2}^{z}\right], \\ \nonumber
\end{eqnarray}
where $J_{0}(x)$ is the Bessel function of the first kind, $a(\lambda_{1}, \lambda_{2}) = \frac{1}{2} \left[J_{0}\left(2(\lambda_{1} - \lambda_{2})\right) + J_{0}(2(\lambda_{1}+\lambda_{2})) \right]$,  $b(\lambda_{1},\lambda_{2})  = J_{0}(2(\lambda_{1}-\lambda_{2})) - a(\lambda_{1}, \lambda_{2})$, $c(\lambda)=\frac{1}{2}\left[1+J_{0}(4\lambda)\right]$, and $d(\lambda)=1-c(\lambda)$.  We have absorbed a factor of $\frac{J_{0}(2\lambda_{1})}{a(\lambda_{1},\lambda_{2})}$ in the definitions of $h_{1}$ and $h_{N}$ \footnote{Note that for finite frequency drives, the $n^{\mathrm{th}}$ order perturbative correction to $H_{\mathrm{F}}^{(0)}$ is of order $1/\omega^{n}$.}. 

A few remarks are in order. First, the periodic driving, $f(t)$, effectively generates multispin interactions [Eqn.~(\ref{eq:stroboscopicHamiltonian})]~\cite{Iadecola_stroboscopic_2015}. Second, while $H_{\mathrm{F}}^{(0)}$ exhibits a $\mathbb{Z}_{2} \times \mathbb{Z}_{2}$ symmetry, the parent Hamiltonian [Eqn.~(\ref{eq:drivenHamiltonian})] possesses only a smaller  $\mathbb{Z}_{2}$  symmetry group, indicating that the ``emergent'' dihedral symmetry of $H_{\mathrm{F}}^{(0)}$ must be broken at higher orders in the Magnus expansion~\cite{suppinfo}.  Finally, the $V_{x}=0$ limit of Eqn.~(\ref{eq:stroboscopicHamiltonian}) describes a pair of decoupled 1D $p$-wave superconductors~\cite{kitaev_pwave} and harbors two simple limits: for $a(\lambda_{1},\lambda_{2}) > b(\lambda_{1}, \lambda_{2})$, the ground state is a trivial insulator, while for $a(\lambda_{1},\lambda_{2})< b(\lambda_{1},\lambda_{2})$, the ground state is a bosonic SPT insulator. The key signature of this latter ESPT phase is the existence of protected modes localized around the boundary of the system. Crucially, the $\lambda_{1},\lambda_{2}$-dimerization of the Ising interaction enables us to arbitrarily tune the correlation length of the edge mode (inset of Fig.~\ref{fig:ESPT}a), leading to coherent dynamics with significantly higher fidelity than those of the undimerized TFIM~\cite{Iadecola_stroboscopic_2015}.

To characterize the edge coherence, we introduce the trace fidelity
$F^{\alpha}(t) = \frac{1}{Z}\mathrm{Tr} \left[e^{-\beta H(t)} \Sigma^{\alpha}(t)\Sigma^{\alpha}(0)\right]$
as a function of time, where $Z$ is the partition function, $\beta = 1/k_{\mathrm{B}}T$, and $\Sigma^{\alpha}$ are the zero correlation length edge operators $\Sigma^{x} = \sigma_{1}^{x}\sigma_{2}^{z}$, $\Sigma^{y} = \sigma_{1}^{y}\sigma_{2}^{z}$, and $\Sigma^{z} = \sigma_{1}^{z}$. This autocorrelation function at infinite temperature will serve as a proxy for the coherence time. Furthermore, since we are interested in coherent MBL-protected dynamics at finite energy densities, from hereon we add strong disorder to the system via random on-site fields $h_{i}$ \footnote{The fields for the ESPT are sampled from uniform distributions with $\averag{h_{\mathrm{bulk}}} = 1.0$ and width $\delta h_{\mathrm{bulk}} = 2.0$ in the bulk ($1<i<N$); $\averag{h_{\mathrm{edge}}} = a(\lambda_{1},\lambda_{2})/J_{0}(2\lambda_{1}) \approx -0.11$ and width $\delta h_{\mathrm{edge}} = 2\averag{h_{\mathrm{edge}}}$ on the edges}. 

As alluded to above, there are two mechanisms of edge spin decoherence introduced by interactions: (1) scattering with thermal excitations and (2) breaking of the $\mathbb{Z}_{2} \times \mathbb{Z}_{2}$ symmetry. While the first is ameliorated via MBL (Fig.~\ref{fig:ESPT}a), the second is intrinsic to the stroboscopic approach---the ESPT phase is stable only up to a finite parametric time scale, $T_{2,\mathrm{symm}}^{*} \sim (h^2/\omega)^{-1}$, beyond which the protecting symmetry is broken.
 
The first effect is reminiscent of similar discussions in the static context~\cite{bahri_localization_2015,sondhi_mbl_order, yao_many-body_2015}, where disorder can localize thermal bulk excitations and suppress scattering. Since the edge operators are odd under the $\mathbb{Z}_{2} \times \mathbb{Z}_{2}$ symmetry, their dressed MBL counterparts will not appear in the effective ``l-bit'' Hamiltonian~\cite{nandkishore_many-body_2015,altman_many-body_2015} and dephasing occurs solely via coupling to the other edge mode~\cite{yao_many-body_2015} on a time scale that is exponential in system size, $T_{2, \mathrm{MBL}}^{*} \sim e^{\mathcal{O}(N)}$\footnote{For each disorder realization, we numerically obtain $F^{\alpha}(t)$, fit an exponential $e^{-t/T_{2}^{*}}$ through the peaks, extract $T_{2}^{*}$, and average over 30-1000 realizations.}, as depicted in Fig.~\ref{fig:ESPT}b. Thus, so long as the effective dynamics are described by $H^{(0)}_{\mathrm{F}}$, one finds that even in the interacting, periodically driven system, disorder can lead to a revival of the coherence time (Fig.~\ref{fig:ESPT}a).

This MBL enhancement of edge coherence is cut off by the fact that the first order Magnus correction, $H_{\mathrm{F}}^{(1)}$, breaks the $\mathbb{Z}_{2} \times \mathbb{Z}_{2}$ symmetry. For time scales $t > T_{2,\mathrm{symm}}^{*}$,  even though bulk excitations remain many-body localized, there is no symmetry protecting the edge operators, which can then scatter locally. Thus, for a finite size system, decoherence in the presence of interactions that preserve the dihedral symmetry occurs on a time scale $T_{2}^{*} \sim \min(T_{2,\mathrm{MBL}}^{*}, T_{2,\mathrm{symm}}^{*}) \sim \min(e^{\mathcal{O}(N)}, \mathcal{O}(\omega/h^2))$ as illustrated in Fig.~\ref{fig:ESPT}b.

\begin{figure}
\centering
\includegraphics[width=1.0\columnwidth]{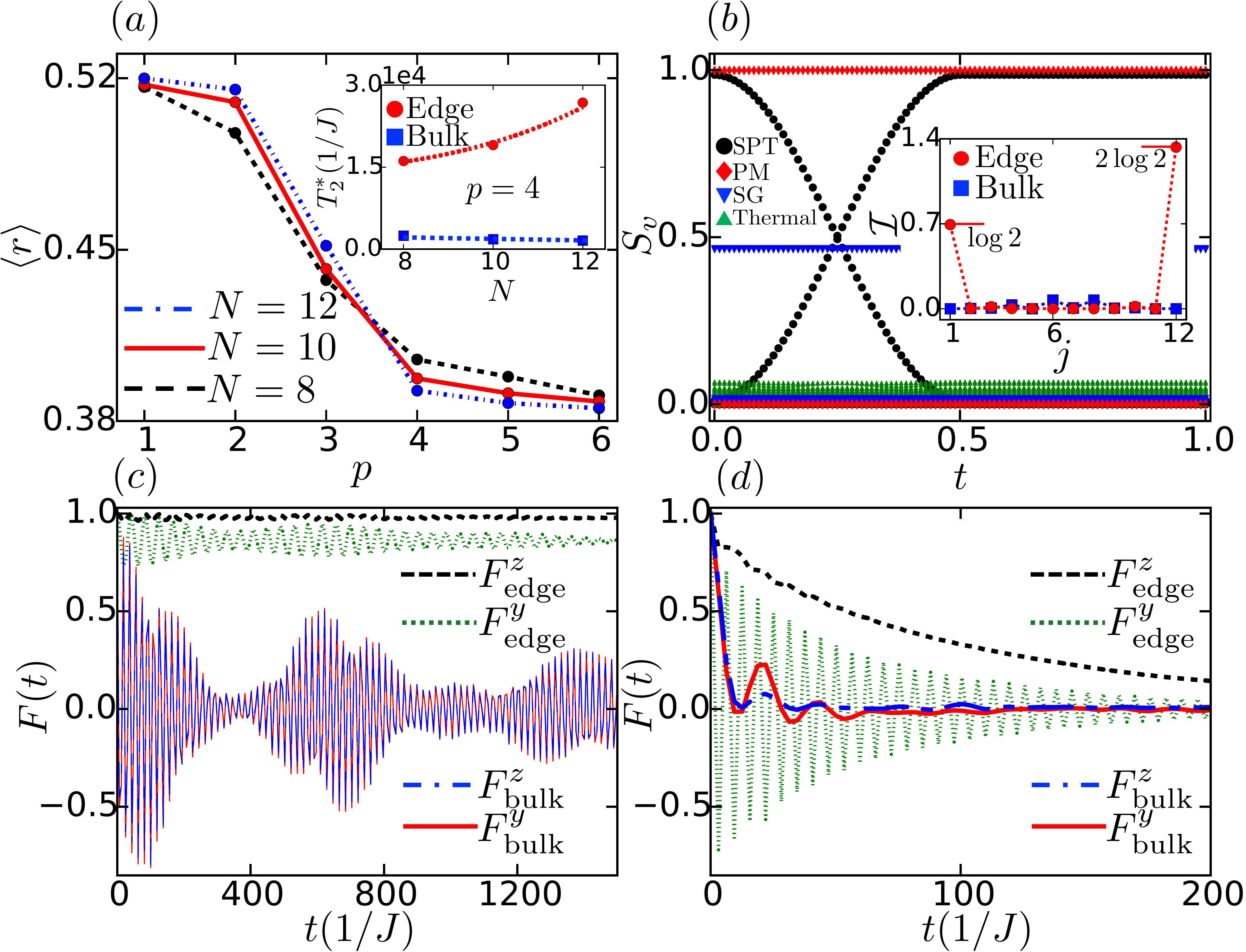}
 \caption[caption]{FSPT phase---(a) The $\langle r\rangle$ ratio as a function of the power law exponent $p$ for a chain with periodic boundary conditions. The $h_{i}$'s are sampled from the uniform distribution $[0.1, 0.9]$ and $T=\pi$ (in units of $J=1$). There is a MBL-delocalization phase transition with a critical point $p_{c}\approx 3.5$. (inset) $T_{2}^{*}$ as a function of $N$, where the edge coherence is fit to $\sim N^4$. 
(b) The entanglement spectrum micromotion for $N=12$. The parameters $(p,T,J,W)$ are: $(4,\pi,1,1)$ for the SPT; $(1,\pi,1,1)$ for the thermal behavior; $(4,\pi,0.05,0.8)$ for the paramagnet; $p=4$, $T=\pi$, $J=0.5$, $h\in[0.5,1]$ for the spin glass. (inset) Mutual information $\mathcal{I}(i,j)=S_{i}+ S_{j}-S_{ij}$ (where $S$ is the von Neumann entropy) within the SPT phase: $\mathcal{I}(1,j)$ (red circles) and $\mathcal{I}(6,j)$ (blue squares).
 (c) $F^{y}(t)$ and $F^{z}(t)$ for the edge and the bulk in a system of $N=10$ spins for the model in Eqn.~(\ref{eq:drivenHamiltonian2}). The bulk curves are almost overlapping.
 (d) Same as in (c), but with an additional term, $V_{x}\sum_{i}\sigma_{i}^{x}\sigma_{i+1}^{x}$ ($V_{x}=0.3$) added to $H_1$.
 }
  \label{fig:FSPT} 
\end{figure}

The addition of a more generic symmetry-breaking interaction term, such as $V_{y}\sum_{i}\sigma_{i}^{y}\sigma_{i+1}^{y}$ or a long-range power-law tail, breaks  the $\mathbb{Z}_{2} \times \mathbb{Z}_{2}$ symmetry  at  lowest order in the Magnus expansion. In this case, there is no parametric time scale where we expect ESPT dynamics (i.e.~$T_{2,\mathrm{symm}}^{*}\sim \mathcal{O}(1)$), and the edge modes rapidly decohere via local scattering (Fig.~\ref{fig:ESPT}b).

\emph{FSPT phase}---To obtain edge modes with coherence that persists to arbitrary times and is robust to long-range interactions, we now turn to the realization of an intrinsically Floquet SPT phase.  We engineer a FSPT phase protected by both  $\mathbb{Z}_2$ symmetry and periodic driving which cannot exist in
equilibrium~\cite{sondhi_floquet_class1,sondhi_floquet_class2, vishwanath_floquet_class, nayak_floquet_class, harper_floquet_class}. Consider the stroboscopic Hamiltonian
\begin{equation}\label{eq:drivenHamiltonian2}
H(t) =  \left\{
      \begin{aligned}
        H_1 = &\sum_{i\neq j} \frac{J}{|R_{i}-R_{j}|^{p}}\sigma_{i}^{z}\sigma_{j}^{z}& \;&\mathrm{if}&\; 0\leq t < T/2\\
        H_2 = &\sum_{i=1}^{N} h_{i}\sigma_{i}^{x}& \;&\mathrm{if}&\; T/2\leq t < T,\\
      \end{aligned}
    \right.
\end{equation}
where $R_{i}=i$ is the position of the $i^{\mathrm{th}}$ spin and $h_{i}\in[0,W]$.  The protecting symmetries are the product of $\sigma_x$ on all sites ($\mathbb{Z}_{2}$) and discrete translations in time ($\mathbb{Z}$).  The unitary evolution under $H(t)$ is given by $U(t) = \mathcal{T} \exp \left(-i\int_{0}^{T} H(t) dt \right)$ and  the Floquet operator by $U = U(T)$. Building upon previous studies~\cite{khemani_phase_2015,jiang_majorana_2011,bastidas_2012,patel_2013}, we expect to observe the FSPT phase at $\frac{JT}{2} \approx \frac{\pi}{2}$~\cite{suppinfo}.

Since the disorder strength is limited to $W \lesssim 1/T$ by the periodic structure of the binary drive~\cite{suppinfo}, the system cannot be localized for arbitrarily strong interactions.  By computing the level-statistics ratio  $\langle r \rangle$ \footnote{For each disorder realization, we diagonalize the Floquet Hamiltonian $H_{\mathrm{F}}$ defined via $U = e^{-iH_{\mathrm{F}}T}$ and obtain a set of quasi-energies $\epsilon_{n}$ modulo $2\pi$. We define the energy gaps as $\delta_{n}= \epsilon_{n+1}-\epsilon_{n}$ and the ratio $r = \min(\delta_{n},\delta_{n+1})/\max(\delta_{n},\delta_{n+1})$. Finally, we average over all the quasi-energies and over 2500-100000 disorder realizations.} as a function of the power-law exponent $p$ (Fig.~\ref{fig:FSPT}a), we observe a clear MBL-delocalization phase transition at $p_{c} \approx 3.5$~\cite{yao2014many}.  For the remainder of the text, we set $p=4$ as a computationally tractable model within the MBL phase.

To probe the nature of edge coherence in the FSPT phase, we again compute the trace fidelity $F^{\alpha}=\frac{1}{2^{N}}\mathrm{Tr} \left[\sigma_{i}^{\alpha}(t)\sigma_{i}^{\alpha}(0)\right]$. As depicted in the inset of Fig.~\ref{fig:FSPT}a, and similar to the ESPT phase, the edge spin exhibits a significantly longer coherence time than bulk spins. However, a crucial difference emerges in the scaling with $N$. For long-range interactions, the coherence time of the ESPT phase scales independently of the system size, $T_{2}^{*} \sim \mathcal{O}(1)$, whereas the FSPT exhibits a quartic scaling $T_{2}^{*} \sim \mathcal{O}(N^4)$ (owing to the $1/R^4$ power-law interactions between the two edge modes), as shown in the inset of Fig.~\ref{fig:FSPT}a).

To further distinguish between the topological features of the ESPT and FSPT phases, we introduce a novel micromotion-based entanglement spectrum signature of the latter~\cite{vishwanath_floquet_class}. In particular, for an eigenstate $\ket{\psi}$ of the Floquet operator $U$, we compute the entanglement spectrum, $\{\eta_{i}(t)\}$, associated with the half-chain cut of $\ket{\psi(t)} = U(t)\ket{\psi}$  for  $0\leq t \leq T$. By Schmidt decomposing $\ket{\psi(t)} = \sum_{i=1}^{2^{N/2}} \eta_{i}(t)\ket{\mathrm{Left}_{i}(t)}\otimes \ket{\mathrm{Right}_{i}(t)}$, we obtain  $\{\eta_{i}(t)\}$ across the two sets, $\{\ket{\mathrm{Left}_{i}(t)}\}$ and $\{\ket{\mathrm{Right}_{i}(t)}\}$, which span the Hilbert spaces of the left and right halves of the chain. Unlike in equilibrium, where a single snapshot of the entanglement spectrum shows the existence of topological edge modes, we find that, at any given time $t$, the spectrum is trivial and there is no signature of FSPT order (Fig.~\ref{fig:FSPT}b). However, by following the micromotion evolution of the spectrum over a single Floquet period, we can robustly identify the topological signature of the FSPT phase~\cite{vishwanath_floquet_class}. 

To see this, we note that the entanglement spectrum is gapped at $t=0$ and $t=T$ which allows us to associate an SPT invariant to each nontrivial band---namely, the $\mathbb{Z}_{2}$ symmetry charge of the corresponding Schmidt states, $\bra{\mathrm{Left}_{i}(t)}\prod_{j} \sigma_{j}^{x}\ket{\mathrm{Left}_{i}(t)}=\pm 1$. There exists a band crossing during the micromotion (Fig.~\ref{fig:FSPT}b), pointing to the fact that the charges of each band are flipping during a Floquet period. This difference between the initial and final $\mathbb{Z}_{2}$ charges cannot be altered without closing the entanglement gap, suggesting that the band crossing is, in fact, a robust feature of FSPT order. Indeed, this nontrivial behavior is absent in the paramagnetic and spin glass phases (Fig.~\ref{fig:FSPT}b). 

Finally, an additional entanglement-based feature of the FSPT's non-trivial protected edge modes is captured by the spatial dependence of two-spin mutual information. We observe a $\log2$ entropy in each edge spin, $2\log2$ mutual information shared between the two edges, and approximately zero mutual information shared between bulk spins (inset of Fig.~\ref{fig:FSPT}b). In combination, this points to the fact that the two edge modes are well localized to a single site and behave like an EPR pair.

\emph{Experimental realization}---Both the ESPT and FSPT Hamiltonians can be implemented in a chain of Rydberg-dressed alkali-metal atoms~\cite{vanBijnen15,Glaetzle15,Goldschmidt16,Zeiher16,zeiher17} trapped in a 1D optical lattice or tweezer array~\cite{Lester15,endres2016cold} (Fig.~\ref{fig:schematic}). The spin degree of freedom is formed by two ground hyperfine states, with a resonant Raman coupling of spatially varying Rabi frequency $h_i$ simulating the on-site transverse fields. Random fields can be formed by optical speckle disorder or with a spatial light modulator.

Strong spin-spin interactions are introduced by coupling state $\ket{\up}$ to a Rydberg state $\ket{\Ryd}$ with an off-resonant laser field of Rabi frequency $\Omega$ and detuning $\Delta>\Omega$. The result is an effective (dressed) Ising interaction~\cite{Gil14,Zeiher16}
\begin{equation}\label{eq:Hdress}
H_I = -\frac{\Omega^4}{8\Delta^3}\frac{1}{1+\abs{R_i-R_j}^6/R_c^6}\sigma_{i}^{z}\sigma_{j}^{z},
\end{equation}
where the interaction range $R_c = \left(-C_6/\Delta\right)^{1/6}$ depends on the van der Waals coefficient $C_6$ of the Rydberg-Rydberg interaction and is typically on the few-micron scale.  At fixed lattice spacing $a_1$, the ratio of nearest to next-nearest-neighbor couplings is set by $R_c$ (Fig.~\ref{fig:schematic}).

While the Rydberg dressing is subject to dissipation from the finite lifetime $\Gamma^{-1}$ of the Rydberg state~\cite{Goldschmidt16,Zeiher16}, the interaction-to-decay ratio can be large~\cite{Glaetzle15,vanBijnen15} in a 1D system.  At fixed Rabi frequency $\Omega$, the ratio of the Ising coupling $J$ to the lifetime $\gamma = (\Omega^2/4\Delta^2)\Gamma$ of the Rydberg-dressed state is limited to $J/\gamma = \frac{\Omega^2}{2\Delta\Gamma} < \frac{\Omega}{\Gamma}$.  This limit is set by the condition $\Omega^2/\Delta^2\ll 1$ that the Rydberg-state population within the radius $R_c\sim a_1$ be small, so that the perturbative result of Eqn.~(\ref{eq:Hdress}) holds.  At realistic laser power on the $6S_{1/2}\rightarrow nP_{3/2}$ transitions (with $n \gtrsim 40$) in cesium~\cite{lee16}, parameters $(\Omega, \Gamma)\approx2\pi\times(4, 0.002)$~MHz allow for large coupling-to-decay ratios $J/\gamma \lesssim 10^3$.

To observe the FSPT phase, we envision initializing the system in a product state with high energy density and letting it undergo unitary time evolution. After each Floquet period $T$, one measures the spin-spin autocorrelation function $\langle \sigma^{\alpha}(nT)\sigma^{\alpha}(0) \rangle$ for both an edge and bulk spin. Numerics (Fig.~\ref{fig:FSPT}c) for $N=10$ atoms indicate that a time $t\sim 10^2/J$ suffices to observe a significant difference between the bulk- and edge-spin fidelities.  The difference can be observed over an even shorter time scale $t \sim 30/J$ (Fig.~\ref{fig:FSPT}d) by adding a decohering interaction term $V_{x}\sum_{i}\sigma_{i}^{x}\sigma_{i+1}^{x}$ to $H_1$ in Eqn.~(\ref{eq:drivenHamiltonian2}).  Experimentally, $V_{x}$ can be introduced by simultaneously dressing both states $\ket{\down}$ and $\ket{\up}$~\cite{Glaetzle15} to generate flip-flop processes $\propto \sigma_{i}^{+}\sigma_{i+1}^{-}$.

To experimentally verify the distinct advantages of the intrinsically Floquet SPT phase, our scheme can be modified to emulate the ESPT phase for comparison.  Realizing the ESPT Hamiltonian requires alternating stroboscopically between ferromagnetic and antiferromagnetic Ising interactions  by simultaneously changing the signs of the detuning $\Delta$ and of the van der Waals coefficient $C_6$.  While a conceptually simple approach is to switch between two different laser fields detuned by $\Delta_{2} \approx -\Delta_{1}$ from two different Rydberg states $\ket{\Ryd_2}, \ket{\Ryd_1}$, a more practical approach may be to dynamically control the sign of $C_6$ with an electric field~\cite{Vogt06}.  We detail concrete level schemes for an implementation in cesium in Ref.~\cite{suppinfo}.

Our proposal raises the tantalizing possibility of observing coherent quantum dynamics at high temperatures in strongly interacting disordered systems~\cite{bahri_localization_2015, sondhi_mbl_order, yao_many-body_2015}. We have studied two different routes towards SPT phases in driven, disordered spin chains: by engineering effective three-spin interactions (ESPT) or by intrinsically dynamical quantized pumping of spin (FSPT).  In both cases, decoherence can be caused by breaking the protecting symmetry group.  However, as the ESPT relies on a symmetry that is only approximately realized in the high frequency limit, it survives only up to a finite time scale for short-range interactions, and is fragile to generic interactions.  By contrast, the FSPT survives at arbitrary times and is robust to long-range interactions.

\emph{Acknowledgements}---We are grateful to Ehud Altman, Alexey Gorshkov, Chris Laumann, Samuel Lederer, and Johannes Zeiher for insightful discussions, and to Tori Borish and Ognjen Markovic for feedback on the manuscript. This work was supported, in part, by  the Miller Institute for Basic Research in Science, AFOSR MURI Grant No. FA9550-14-1-0035, the NSF PHY-1654740, the Simons Investigator Program, the Gordon and Betty Moore Foundation's EPiQS Initiative through Grant GBMF4307, the ARO, and the Alfred P. Sloan Foundation.

\bibliography{SPT_biblio}

\end{document}

\title{Supplementary Material: Floquet Symmetry-Protected Topological Phases in Cold Atomic Systems}
\author{I.-D. Potirniche, A. C. Potter, M. Schleier-Smith, A. Vishwanath, and N. Y. Yao}
\maketitle

\section{The Floquet Hamiltonian for the ESPT}

We begin by providing details about the Magnus expansion and deriving Eqn.(2) from the main text. First, we define a unitary transformation to a ``rotating frame'', $U_{\mathrm{R}}(t) = \exp\left(i \sin(\omega t) \sum_{i=1}^{N-1} \lambda_{i}\sigma_{i}^{z}\sigma_{i+1}^{z}\right)$, which maps an arbitrary state $\ket{\psi(t)}$ to $\ket{\psi_{\mathrm{R}}(t)} = U_{\mathrm{R}}(t) \ket{\psi(t)}$. At stroboscopic times $t_{n} = n\frac{2\pi}{\omega} = n T$ ($n \in \mathbb{Z}$), one finds that $\ket{\psi_{\mathrm{R}}(t_{n})} = e^{-iH_{\mathrm{F}}t_{n}}\ket{\psi_{\mathrm{R}}(0)}$, where $H_{\mathrm{F}}$ is the so-called Floquet Hamiltonian which we construct perturbatively in orders of $1/\omega$; the Schrodinger equation becomes $i \partial_{t} \ket{\psi_{\mathrm{R}}(t)} = H_{\mathrm{R}}(t) \ket{\psi_{\mathrm{R}}(t)}$, where the ``rotated'' Hamiltonian is $H_{\mathrm{R}}(t) =   U_{\mathrm{R}}(t) H(t)U_{\mathrm{R}}^{\dagger}(t) - iU_{\mathrm{R}}(t)\partial_{t}U_{\mathrm{R}}^{\dagger}(t)$. Using the driven TFIM Hamiltonian from Eqn.(1) in the main text, we obtain
\begin{equation*}
H_{\mathrm{R}}(t) = \sum_{i=1}^{N}h_{i}U_{\mathrm{R}}\sigma_{i}^{x}U_{\mathrm{R}}^{\dagger} + V_{x}\sum_{i=1}^{N-1}U_{\mathrm{R}}\sigma_{i}^{x}U_{\mathrm{R}}^{\dagger}U_{\mathrm{R}}\sigma_{i+1}^{x}U_{\mathrm{R}}^{\dagger}.
\end{equation*}
From the explicit form for $U_{\mathrm{R}}(t)$ we immediately find
\begin{eqnarray}\label{eq:rotated_x}
U_{\mathrm{R}}(t)\sigma_{i}^{x}U_{\mathrm{R}}^{\dagger}(t) &=& \sigma_{i}^{x} \left[\cos(2\tilde{\lambda}_{i-1}(t)) - i\sin(2\tilde{\lambda}_{i-1}(t))\sigma_{i-1}^{z}\sigma_{i}^{z} \right] \left[\cos(2\tilde{\lambda}_{i}(t)) - i\sin(2\tilde{\lambda}_{i}(t))\sigma_{i}^{z}\sigma_{i+1}^{z} \right] \\ \nonumber
U_{\mathrm{R}}(t)\sigma_{i}^{x}\sigma_{i+1}^{x}U_{\mathrm{R}}^{\dagger}(t) &=& \sigma_{i}^{x}\sigma_{i+1}^{x} \left[\cos(2\tilde{\lambda}_{i-1}(t)) - i\sin(2\tilde{\lambda}_{i-1}(t))\sigma_{i-1}^{z}\sigma_{i}^{z} \right] \left[\cos(2\tilde{\lambda}_{i+1}(t)) - i\sin(2\tilde{\lambda}_{i+1}(t))\sigma_{i+1}^{z}\sigma_{i+2}^{z} \right], \\ \nonumber
\end{eqnarray}
where $\tilde{\lambda}(t) = \lambda\sin(\omega t)$. For the operators acting at the boundary, we define $\lambda_{0}= \lambda_{N}=0$.

\textbf{Lowest order term:} The lowest order term in the Magnus expansion for the Floquet Hamiltonian is just the time-averaged rotated Hamiltonian: $H_{\mathrm{F}}^{(0)} = \frac{1}{T}\int_{0}^{T}H_{\mathrm{R}}(t')dt'$, where $T = \frac{2\pi}{\omega}$ is the period of driving. To avoid cluttering our formulae, we use $\frac{1}{T}\int_{0}^{T}dt (...) \equiv \langle...\rangle$. 

Then we immediately obtain these identities: $\left\langle\cos(2\tilde{\lambda}(t))\right\rangle = J_{0}(2\lambda)$ and $\left\langle\sin(2\tilde{\lambda}(t))\right\rangle = 0$. With these in hand, we also find
\begin{eqnarray}
\left\langle \cos(2\tilde{\lambda}_1(t)) \sin(2 \tilde{\lambda}_{2}(t)) \right\rangle &=& 0 \\ \nonumber
\left\langle \cos(2\tilde{\lambda}_{1}(t)) \cos(2\tilde{\lambda}_{2}(t)) \right\rangle &=& a(\lambda_{1},\lambda_{2}) \\ \nonumber
\left\langle \sin(2\tilde{\lambda}_{1}(t)) \sin(2\tilde{\lambda}_{2}(t))\right\rangle &=& b(\lambda_{1},\lambda_{2}), \\ \nonumber
\end{eqnarray}
where $a(\lambda_1,\lambda_2) = \frac{1}{2}\left[J_{0}(2(\lambda_1-\lambda_2))+J_{0}(2(\lambda_1+\lambda_2))\right]$ and $b(\lambda_1,\lambda_2) = J_{0}(2(\lambda_1-\lambda_2)) - a(\lambda_1,\lambda_2)$. The ratio $b(\lambda_1,\lambda_2)/a(\lambda_1,\lambda_2)$ controls the correlation length of the edge mode and the $\lambda_1,\lambda_2$ -dimerization enables us to arbitrarily tune it (Fig.~\ref{fig:exp_enhancement}b).

Expanding the terms from Eqn.~\ref{eq:rotated_x} and applying these identities, we get
\begin{eqnarray}\label{eq:lowestorder}
\left\langle\sum_{i=1}^{N}h_{i}U_{\mathrm{R}}\sigma_{i}^{x}U_{\mathrm{R}}^{\dagger}\right\rangle &=& J_{0}(2\lambda_{1})\left(h_{1}\sigma_{1}^{x}+h_{N}\sigma_{N}^{x}\right) + \sum_{i=2}^{N-1}h_{i}\left[a(\lambda_{1},\lambda_{2})\sigma_{i}^{x} -b(\lambda_{1},\lambda_{2})\sigma_{i-1}^{z}\sigma_{i}^{x}\sigma_{i+1}^{z}\right] \\ \nonumber
\left\langle\sum_{i=1}^{N-1}V_{x}U_{\mathrm{R}}\sigma_{i}^{x}\sigma_{i+1}^{x}U_{\mathrm{R}}^{\dagger}\right\rangle &=& V_{x}J_{0}(2\lambda_{2})\left(\sigma_{1}^{x}\sigma_{2}^{x}+ \sigma_{N-1}^{x}\sigma_{N}^{x}\right) + \sum_{i=2}^{N-2}V_{x}\left[c(\lambda_{i+1})\sigma_{i}^{x}\sigma_{i+1}^{x} + d(\lambda_{i+1})\sigma_{i-1}^{z}\sigma_{i}^{y}\sigma_{i+1}^{y}\sigma_{i+2}^{z}\right],
\end{eqnarray}
which corresponds to the expression for $H_{\mathrm{F}}^{(0)}$ in the main text. Note that all terms commute with $\Theta_{\mathrm{even}}=\prod_{i}\sigma_{2i}^{x}$ and $\Theta_{\mathrm{odd}}=\prod_{i}\sigma_{2i+1}^{x}$ which generate the $\mathbb{Z}_{2}\times\mathbb{Z}_{2}$ symmetry.

If we add another term, $V_{y}\sum_{i=1}^{N-1}\sigma_{i}^{y}\sigma_{i+1}^{y}$ to the driven TFIM Hamiltonian, then its contribution to $H_{\mathrm{F}}^{(0)}$ would be identical to the one on the second line of Eqn.~\ref{eq:lowestorder} with $[x\rightarrow y]$. In that case, we note that the Floquet Hamiltonian would have only a smaller, $\mathbb{Z}_{2}$, symmetry group.

\textbf{First order correction:} The first order correction in the Magnus expansion is 
\begin{equation}\label{eq:first_order_correction}
H_{\mathrm{F}}^{(1)} = -\frac{i}{2T} \int_{0}^{T}dt_{1} \int_{0}^{t_{1}}dt_{2}  \left[ H_{\mathrm{R}}(t_{1}), H_{\mathrm{R}}(t_{2}) \right].
\end{equation}
We want to show that $H_{\mathrm{F}}^{(1)}$ contains a term that breaks the $\mathbb{Z}_{2}\times\mathbb{Z}_{2}$ symmetry of $H_{\mathrm{F}}^{(0)}$. More specifically, there exists a contribution of the form $\sum_{i}f_{i}\sigma_{i}^{y}\sigma_{i+1}^{y}$, where $f_{i} \sim  \mathcal{O}(\frac{h^{2}}{\omega})$. For simplicity, we will show that this holds even in the non-interacting case, $V_{x} = V_{y} = 0$.

Neglecting terms acting at the boundary, from Eqn.~\ref{eq:rotated_x}, we get
\begin{eqnarray*}
H_{\mathrm{R}}(t) &=& \sum_{i} h_{i}\left[ \cos(2\tilde{\lambda}_{1}(t))\cos(2\tilde{\lambda}_{2}(t)) \sigma_{i}^{x} - \cos(2\tilde{\lambda}_{i-1}(t))\sin(2\tilde{\lambda}_{i}(t))\sigma_{i}^{y}\sigma_{i+1}^{z}- \cos(2\tilde{\lambda}_{i}(t))\sin(2\tilde{\lambda}_{i-1}(t))\sigma_{i-1}^{z}\sigma_{i}^{y}\right] \\
 &-&\sum_{i} h_{i}\sin(2\tilde{\lambda}_{1}(t))\sin(2\tilde{\lambda}_{2}(t))\sigma_{i-1}^{z}\sigma_{i}^{x}\sigma_{i+1}^{z}.
\end{eqnarray*}

When calculating the commutator $\left[H_{\mathrm{R}}(t_{1}),H_{\mathrm{R}}(t_{2})\right]$, the only terms that give a $\sigma_{i}^{y}\sigma_{i+1}^{y}$ (or YY) contribution are
\begin{eqnarray*}
&i&\sum_{i} 2h_{i-1}h_{i} \left[\cos(2\tilde{\lambda}_{1}(t_1))\cos(2\tilde{\lambda}_{2}(t_1))\cos(2\tilde{\lambda}_{i}(t_{2}))\sin(2\tilde{\lambda}_{i-1}(t_2))\right] \sigma_{i-1}^{y}\sigma_{i}^{y} \\
-&i&\sum_{i}2h_{i-1}h_{i}\left[ \cos(2\tilde{\lambda}_{1}(t_2))\cos(2\tilde{\lambda}_{2}(t_2))\cos(2\tilde{\lambda}_{i}(t_{1}))\sin(2\tilde{\lambda}_{i-1}(t_1))\right]\sigma_{i-1}^{y}\sigma_{i}^{y} \\
&i& \sum_{i}2h_{i}h_{i+1}\left[\cos(2\tilde{\lambda}_{1}(t_1))\cos(2\tilde{\lambda}_{2}(t_1))\cos(2\tilde{\lambda}_{i-1}(t_{2}))\sin(2\tilde{\lambda}_{i}(t_{2})) \right]\sigma_{i}^{y}\sigma_{i+1}^{y} \\
-&i& \sum_{i}2h_{i}h_{i+1}\left[\cos(2\tilde{\lambda}_{1}(t_2))\cos(2\tilde{\lambda}_{2}(t_2))\cos(2\tilde{\lambda}_{i-1}(t_{1}))\sin(2\tilde{\lambda}_{i}(t_{1})) \right]\sigma_{i}^{y}\sigma_{i+1}^{y}.
\end{eqnarray*}

Combining all of these terms together, we obtain 
\begin{equation}
i\sum_{i}4h_{i-1}h_{i} \left[\cos(2\tilde{\lambda}_{1}(t_1))\cos(2\tilde{\lambda}_{2}(t_1))\cos(2\tilde{\lambda}_{i}(t_{2}))\sin(2\tilde{\lambda}_{i-1}(t_2))- \left(t_{1}\leftrightarrow t_{2}\right) \right] \sigma_{i-1}^{y}\sigma_{i}^{y}. 
\end{equation}
Plugging this expression into Eqn.~\ref{eq:first_order_correction}, we finally see that $H_{\mathrm{F}}^{(1)}$ contains a term of the form $\sum_{i}\frac{h_{i-1}h_{i}}{\omega}f_{i}\sigma_{i-1}^{y}\sigma_{i}^{y}$, where
\begin{equation}
f_{i} = \frac{1}{\pi}\int_{0}^{2\pi}du_{1}\int_{0}^{u_{1}}du_{2}\cos(2\lambda_{i}\sin{u_{1}})\cos(2\lambda_{i}\sin{u_{2}})\sin[2\lambda_{i-1}(\sin{u_{2}} - \sin{u_{1}})].
\end{equation}
For the dimerized couplings used in the main text, $\lambda_{1} = 1.54$ and $\lambda_{2} = 0.69$, we obtain $f_{1} \approx -0.08$ and $f_{2}\approx 0.08$. Thus, the ESPT can only exist for a finite time scale $T_{2,\mathrm{symm}}^{*} \sim (h^2/\omega)^{-1}$ in the thermodynamic limit. At later times, the symmetry of the Floquet Hamiltonian is $\mathbb{Z}_{2}$ which cannot support SPT order.

\section{Decoherence in the ESPT}

\begin{table}[h]
\refstepcounter{table}\label{table:table1}
\begin{tabular}{|l|l|l|l|}
\hline
Interactions & Symmetry & In Majorana operators& $T_{2}^{*}$ \\ \hline
XX ($V_{y} = 0$) & $\mathbb{Z}_{2}\times \mathbb{Z}_{2}$ & Quartic & $\min(\mathcal{O}(\omega), e^{\mathcal{O}(N)})$\\ \hline
YY ($V_{x} = 0$) & $\mathbb{Z}_{2}$ & Quadratic & $\infty$\\ \hline
XY ($V_{y} = 4V_{x}$) & $\mathbb{Z}_{2}$ & Quartic & $\mathcal{O}(1)$\\ \hline
\end{tabular}
\end{table}

\begin{figure}[H]
\centering
\includegraphics[width=\columnwidth]{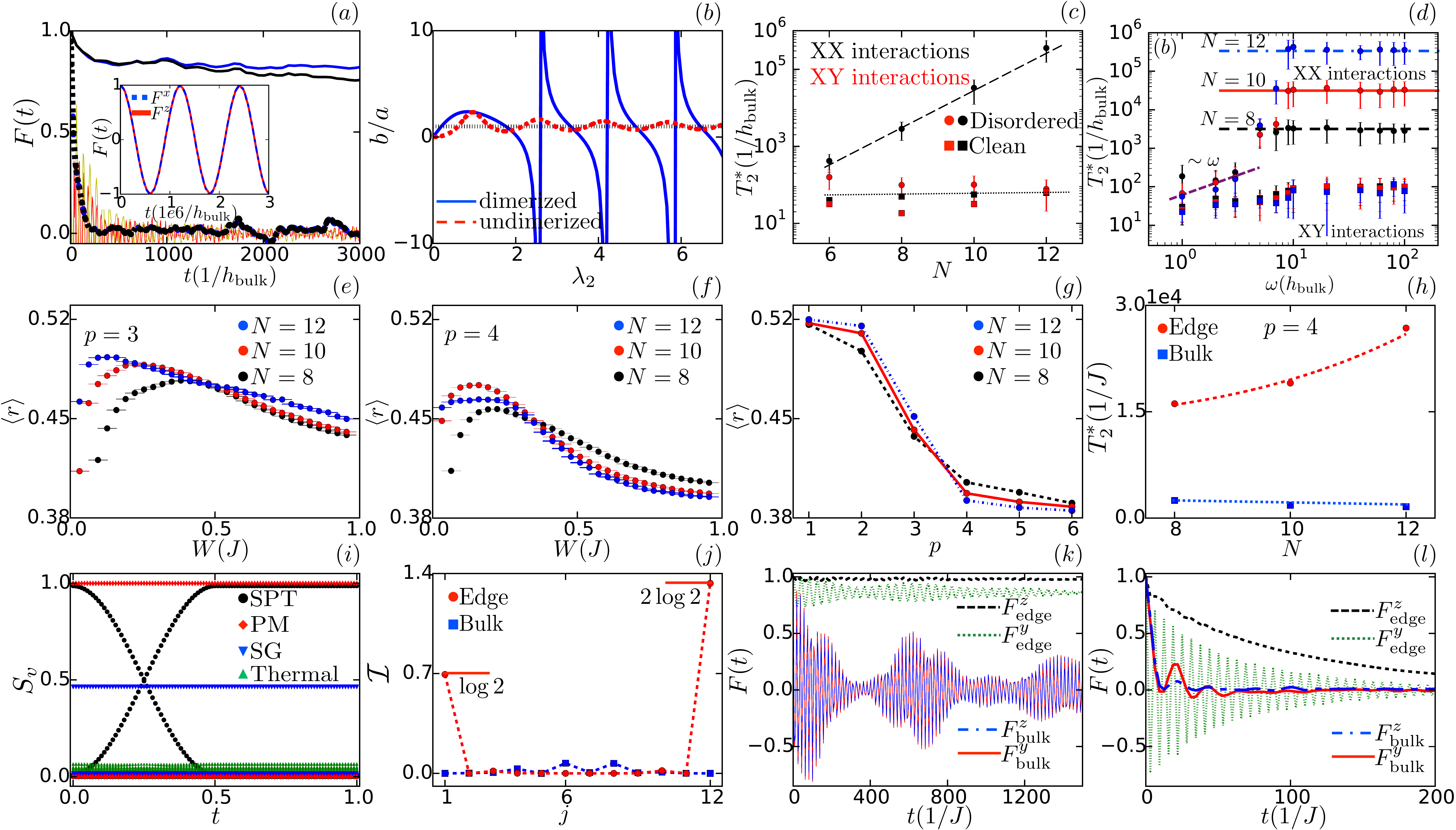}
\caption{(a)-(d): The ESPT described by the Hamiltonian in Eq.(1) from the main text. \\
(a) The overlapping dashed lines (black and blue for $F^{z}$ and $F^{x}$, respectively) correspond to the interacting and clean (uniform fields $h_i$) ESPT. We see that interactions ($V_{x}\neq 0$) lead to decoherence.
The solid black and blue lines (for $F^{z}$ and $F^{x}$, respectively) correspond to the interacting ESPT in the presence of strongly disordered on-site fields $h_i$. We see that disorder leading to MBL restores the coherence in the \emph{dimerized} case ($\lambda_{1} = 1.54$ and $\lambda_{2} =0.69$).
The thin solid yellow and red lines (for $F^{z}$ and $F^{x}$, respectively) correspond to the interacting and strongly-disordered ESPT in the \emph{un-dimerized} case ($\lambda_{1}=\lambda_{2}$). In this model, it is difficult to balance both the localized and topological nature of the phase. The strong disorder needed for MBL leads to large fluctuations of the effective couplings in $H_{\mathrm{F}}^{(0)}$ which introduces trivial puddles that fracture the original topological phase.
(inset) The dashed blue and solid red curves corresponding to $F^{x}$ and $F^{z}$ in the non-interacting ($V_{x} = V_{y} = 0$) and clean ESPT. We observe coherent oscillations whose contrast never decays (due to the lack of interactions) with a period $\sim 10^6$.
(b) Ratio $b(1, \lambda_{2})/a(1, \lambda_{2})$ in the dimerized (solid blue) and the un-dimerized (dashed red) models. The SPT phase corresponds to $b/a > 1$ (dotted black). The dimerized model allows us to arbitrarily tune the correlation length of the edge mode.
(c) $T_{2}^{*}$ as a function of $N$ in both clean (squares) and disordered (circles) systems for the ESPT. The interaction parameters are $V_{x} = 0.05$, $V_{y} = 0$ for XX interactions (black) and $V_{y} = 4V_{x} = 0.2$ for XY interactions (red). The exponential enhancement of the coherence time is seen only in the disordered XX case in which the $\mathbb{Z}_{2} \times \mathbb{Z}_{2}$ symmetry is unbroken in $H_{\mathrm{F}}^{(0)}$.
(d)  $T_{2}^{*}$ as a function of frequency and system size. As $\omega$ is increased for XX interactions (circles), $T_{2}^{*}$ saturates consistent with being bounded by  $T_{2}^{*} \sim \min(\mathcal{O}(\omega), e^{\mathcal{O}(N)})$. Turning on XY interactions (squares), breaks the $\mathbb{Z}_{2} \times \mathbb{Z}_{2}$ symmetry at the level of $H_{\mathrm{F}}^{(0)}$ and leads to a breakdown of the edge coherence for all parameters.\\
\\
(e)-(l): The FSPT described by the Hamiltonian in Eq.(3) from the main text. The on-site fields $h_{i}$ are sampled from the uniform distribution on $[0,W]$.
\\
(e) The level statistics ratio $\langle r \rangle$ as a function of the disorder width $W$ with the interactions power law exponent $p=3$. Maximal disorder width $W=W_{\mathrm{max}}=1$ is not strong enough to localize the system even though $\langle r \rangle$ is below the GOE value of 0.527.
(f) $\langle r \rangle$ as a function of $W$ for $p=4$. At strong disorder strengths, $\langle r \rangle$ is close to the Poisson value of 0.386 signaling localization. At low disorder strengths, there is an anomalous scaling with $N$, but this provides a bound on the  location of the critical point.
(g) The $\langle r\rangle$ ratio as a function of the power law exponent $p$ for the maximal disorder width. There is an MBL-delocalization phase transition with a critical point $p_{c}\approx 3.5$. For the remaining plots, we set $p=4$.
(h) $T_{2}^{*}$ as a function of $N$ for both the edge and bulk fidelities.  The edge coherence is fit to $\sim N^4$ due to the $1/N^4$ power-law interactions between the two edge modes. The bulk coherence does not show signs of enhancement. 
(i) The entanglement spectrum micro-motion. We see that the FSPT exhibits a non-trivial entanglement band-crossing  due to the pumping of a $\mathbb{Z}_2$ symmetry charge, whereas the other phases have a trivial set of bands.
(j) The two-spin mutual information $\mathcal{I}(i,j)=S_{i}+ S_{j}-S_{ij}$ (where $S$ is the von Neumann entropy) within the FSPT phase: $\mathcal{I}(1,j)$ (red circles) and $\mathcal{I}(6,j)$ (blue squares). We observe $\log 2$ entropy in each edge spin, $2\log 2$ mutual information shared between the two edges, and approximately zero (up to finite size effects) mutual information shared between bulk spins. This points to the fact that the two edge modes are well localized to a single site and behave like an EPR pair.
(k) The edge and bulk fidelities of the FSPT for experimentally accessible parameters. The difference between the two can be observed on a time scale $\sim 10^2/J$.
(l) The difference between the edge- and bulk-spin coherence can be observed over an even shorter time scale $\sim 30/J$ by adding a decohering term $V_{x}\sum_{i}\sigma_{i}^x\sigma_{i+1}^x$ to the $H_1$ Hamiltonian in Eq.(3) from the main text.
}
\label{fig:exp_enhancement}
\end{figure}

We now consider a detailed analysis of the interplay between symmetry and decoherence in the ESPT phase. We start from the \emph{non-interacting} driven TFIM [Eqn.(1) from the main text with $V_{x} = 0$]. Coherent oscillations (whose contrast never decays) are observed with a period $\tau = \frac{\pi}{a(\lambda_{1}, \lambda_{2})}\left(\frac{b(\lambda_{1}, \lambda_{2})}{a(\lambda_{1}, \lambda_{2})}\right)^{N/2 - 1}$ [inset of Fig.~\ref{fig:exp_enhancement}a] which can easily be understood by recasting the Hamiltonian using a Jordan-Wigner transformation. Since the original Hamiltonian is quadratic in Majorana operators, and hence non-interacting~\cite{kitaev_pwave,yao_many-body_2015}, the resulting Floquet Hamiltonian must also describe free fermions (at all orders in the Magnus expansion). Thus, even though the first order correction, $H_{\mathrm{F}}^{(1)}$, explicitly breaks the protecting $\mathbb{Z}_{2} \times \mathbb{Z}_{2}$ symmetry as seen in the

\noindent previous section, it does \emph{not} lead to decoherence owing to the lack of scattering between the free fermions. The observed coherent oscillations are simply due to finite-size interactions between the edge spins mediated by their bulk tails.

We now switch to the \emph{interacting} case and consider interactions of the form $H_{\mathrm{int}} = \sum_{i=1}^{N-1}  V_{x} \sigma_{i}^{x}\sigma_{i+1}^{x} + V_{y}\sigma_{i}^{y}\sigma_{i+1}^{y}$. Interestingly, the role of the XX and YY interactions are quite different, as summarized in Table~\ref{table:table1}.  In the limit $V_{x} = 0$, $V_{y} \neq 0$ (YY interactions), $H_{\mathrm{F}}^{(0)}$ breaks the $\mathbb{Z}_{2} \times \mathbb{Z}_{2}$ symmetry. Despite this fact, since the YY interaction preserves the non-interacting nature of the Hamiltonian, we still observe coherent oscillations. Conversely, if $V_{y} = 0$, $V_{x} \neq 0$ (XX interactions), then the stroboscopic Hamiltonian will include quartic terms. In the clean case, this would lead to the immediate decoherence of the edge modes (Fig.~\ref{fig:exp_enhancement}a). However, in the presence of strong disorder, since $H_{\mathrm{F}}^{(0)}$ preserves the $\mathbb{Z}_{2} \times \mathbb{Z}_{2}$ symmetry of the ESPT (Table~\ref{table:table1}), we observe an exponential enhancement of the edge coherence time, $T_{2}^{*}$, compared to the clean case (Fig.~\ref{fig:exp_enhancement}c). We note in passing that in the undimerized case, $\lambda_{1}=\lambda_{2}$, it is difficult to balance both the localized and topological nature of the phase. The strong disorder needed for MBL leads to large fluctuations of the effective couplings in $H_{\mathrm{F}}^{(0)} $ which introduces trivial puddles that fracture the original topological phase (Fig.~\ref{fig:exp_enhancement}a). Finally, if both $V_{x} \neq 0$ and $V_{y} \neq 0$ (XY interactions) the protecting $\mathbb{Z}_{2} \times \mathbb{Z}_{2}$ symmetry is already broken at lowest order in the Magnus expansion, namely $H_{\mathrm{F}}^{(0)}$. In this case, there is no parametric scale where one expects SPT dynamics ($T_{2,\mathrm{symm}}^{*}\sim \mathcal{O}(1)$) and the Hamiltonian is also strongly interacting; thus, the edge modes decohere immediately via local scattering and disorder cannot revive the coherence time compared to the clean case (Fig~\ref{fig:exp_enhancement}c and Fig.~\ref{fig:exp_enhancement}d).

\section{Details on the FSPT}

\subsection{The location and properties of the phase}

In the main text we have mentioned that we expect to observe the FSPT phase at $\frac{JT}{2} \approx \frac{\pi}{2}$. To see why this is the case, let us consider a simplified version of the Hamiltonian defined in Eqn.(3) in the main text:
\begin{equation}\label{eq:drivenHamiltonian2}
H(t) =  \left\{
      \begin{aligned}
        H_1 = &\sum_{i} J\sigma_{i}^{z}\sigma_{i+1}^{z}& \;&\mathrm{if}&\; 0\leq t < T/2\\
        H_2 = &\sum_{i=1}^{N} h_{i}\sigma_{i}^{x}& \;&\mathrm{if}&\; T/2\leq t < T.\\
      \end{aligned}
    \right.
\end{equation}
This model has only nearest-neighbor interactions, but it is exactly solvable and illustrative of the main properties of the FSPT~\cite{khemani_phase_2015,sondhi_floquet_class1,sondhi_floquet_class2}. 

The Floquet operator over a period $T$ can be written as $U = \exp(-iH_{2}T/2)\exp(-iH_{1}T/2)$. Taking $\frac{JT}{2} = \frac{\pi}{2}$ and using open boundary conditions, we find that
\begin{eqnarray*}
\exp\left(-iH_{1}T/2\right) &=& \prod_{i=1}^{N-1} \exp \left(-i\frac{\pi}{2}\sigma_{i}^{z}\sigma_{i+1}^{z}\right) \\
&=& \prod_{i=1}^{N-1} \left[-i\sin\left(\frac{\pi}{2}\right)\sigma_{i}^{z}\sigma_{i+1}^{z} + \cos\left(\frac{\pi}{2}\right)\mathbb{1}\right] \\
&\propto & \prod_{i=1}^{N-1} \sigma_{i}^{z}\sigma_{i+1}^{z} \\
&=& \sigma_{1}^{z}\sigma_{N}^{z}.
\end{eqnarray*}
Note that under periodic boundary conditions this would be proportional to the identity $\mathbb{1}$. 

Secondly, we can also re-write the other piece of the Floquet operator as
\begin{eqnarray*}
\exp\left(-iH_{2}T/2\right) &=& \exp\left(-i\frac{h_1T}{2}\sigma_{1}^x\right)\exp\left(-i\frac{h_N T}{2}\sigma_{N}^{x}\right) \exp\left(-i\sum_{1<i<N} \frac{h_iT}{2}\sigma_{i}^{x}\right) \\
&=& \left[-i\sin\left(\frac{h_1T}{2}\right)\sigma_1^x + \cos\left(\frac{h_1T}{2}\right)\mathbb{1}\right] \left[-i\sin\left(\frac{h_NT}{2}\right)\sigma_N^x + \cos\left(\frac{h_NT}{2}\right)\mathbb{1}\right] \exp\left(-i\sum_{1<i<N} \frac{h_iT}{2}\sigma_{i}^{x}\right) \\
&=& \left[-i\sin(\theta_1)\sigma_{1}^x + \cos(\theta_1)\right] \left[-i\sin(\theta_N)\sigma_{N}^x + \cos(\theta_N)\right]\exp\left(-iH_{\mathrm{bulk}}T/2\right),
\end{eqnarray*}
where $\theta_i = \frac{h_iT}{2}$ and $H_{\mathrm{bulk}} = \sum_{1<i<N} h_i \sigma_{i}^{x}$.

Putting both pieces together, we find
\begin{eqnarray*}
U &\propto & \left[-i\sin(\theta_1)\sigma_{1}^x + \cos(\theta_1)\right] \sigma_{1}^{z}\left[-i\sin(\theta_N)\sigma_{N}^x + \cos(\theta_N)\right]\sigma_{N}^{z}\exp\left(-iH_{\mathrm{bulk}}T/2\right) \\
&\propto & \left[-\sin(\theta_1)\sigma_{1}^y + \cos(\theta_1)\sigma_1^{z}\right] \left[-\sin(\theta_N)\sigma_{N}^y + \cos(\theta_N)\sigma_N^{z}\right]\exp\left(-iH_{\mathrm{bulk}}T/2\right) \\
&\propto & \tilde{\sigma}_{1}^{z}\tilde{\sigma}_{N}^{z}\exp\left(-iH_{\mathrm{bulk}}T/2\right),
\end{eqnarray*}
where $\tilde{\sigma}_{1,N}^{z} = \left[-\sin(\theta_{1,N})\sigma_{1,N}^y + \cos(\theta_{1,N})\sigma_{1,N}^{z}\right]$ and note that $\left(\tilde{\sigma}_{1,N}^{z}\right)^2=1$. Thus, the Floquet operator $U= \tilde{\sigma}_{1}^{z}\tilde{\sigma}_{N}^{z}\exp\left(-iH_{\mathrm{bulk}}T/2\right)$ has two pieces: a ``charge pump'' term $\tilde{\sigma}_{1}^{z}\tilde{\sigma}_{N}^{z}$ corresponding to the product of two operators localized around the edges; and a unitary time evolution over $T/2$ with a local \emph{bulk} Hamiltonian $H_{\mathrm{bulk}} = \sum_{1<i<N} h_i \sigma_{i}^{x}$.

A few observations are in order. The operators $\tilde{\sigma}_{1,N}^{z}$ localized around the edges commute with each other, with the bulk operator $H_{\mathrm{bulk}}$, and, thus, with the Floquet operator $U(T)$. Secondly, these edge operators are odd under the global $\mathbb{Z}_{2}$ symmetry generated by $\Theta = \prod \sigma_i^{x}$:
\begin{eqnarray*}
\Theta \tilde{\sigma}_{1}^{z} \Theta^{\dagger} &=& \sigma_{1}^{x} \left[-\sin(\theta_{1})\sigma_{1}^y + \cos(\theta_{1})\sigma_{1}^{z}\right]\sigma_{1}^{x} \\
&=& \left[\sin(\theta_{1})\sigma_{1}^y - \cos(\theta_{1})\sigma_{1}^{z}\right]\left(\sigma_{1}^{x}\right)^2 \\
&=& -\tilde{\sigma}_{1}^{z},
\end{eqnarray*}
and a similar relation holds for the other edge mode. Lastly, we see that $U(2T) = U^2 = \mathbb{1}_{\mathrm{edges}} \otimes \exp\left(-iH_{\mathrm{bulk}}T\right)$ which means that a generic edge observable, $\mathcal{O}_{\mathrm{edge}}$, satisfies $\mathcal{O}_{\mathrm{edge}}(nT) = \mathcal{O}_{\mathrm{edge}}(nT + 2T)$ for $n\in \mathbb{Z}$, namely it oscillates with a $2T$ period. 

Thus, for $\frac{JT}{2} = \frac{\pi}{2}$ the exactly solvable model with nearest-neighbor interactions has two coherent modes localized around the edges which are odd under the protecting symmetry corresponding to a global spin flip $\prod \sigma_{i}^x$ (i.e. a $\mathbb{Z}_2$ symmetry) and discrete translations in time (i.e. a $\mathbb{Z}$ symmetry). In fact, as argued in~\cite{sondhi_floquet_class1,sondhi_floquet_class2,vishwanath_floquet_class, nayak_floquet_class, harper_floquet_class} and as sketched in the section below, this is true also for the model with generic interactions such as power laws [Eqn.(3) in the main text] in the presence of strong disorder leading to MBL.

\subsection{The role of disorder}

In the FSPT, similar to ESPT phase, disordered on-site fields, $h_{i}\in[0,W]$, play an equally crucial role in restoring the coherence of the edge modes (Fig.~\ref{fig:exp_enhancement}h). Yet their role is even more nuanced. The classification and stability of interacting Floquet-SPT phases of matter~\cite{sondhi_floquet_class1,sondhi_floquet_class2, vishwanath_floquet_class, nayak_floquet_class, harper_floquet_class} hinge on the system exhibiting MBL because this ensures that the eigenstates of the Floquet operator $U$ and Hamiltonian $H_{\mathrm{F}}$ are short-range entangled. The $\mathbb{Z}_{2}$ symmetry corresponding to a global spin flip combined with the $\mathbb{Z}$ symmetry of $H_{\mathrm{F}}$ generated by discrete time translations ensures that the FSPT phase is protected by $\mathbb{Z}\times\mathbb{Z}_{2}$ and has no ground state (static) counterpart. Furthermore, based on the classification of Floquet-MBL-SPT phases via the K\"unneth formula for group cohomology~\cite{sondhi_floquet_class1,sondhi_floquet_class2, vishwanath_floquet_class, nayak_floquet_class, harper_floquet_class}, each application of the Floquet unitary $U$ pumps a lower dimensional SPT through the boundary: the analysis for the exactly solvable model in the above section generalizes to $U = \tilde{\sigma}_{1}\tilde{\sigma}_{N}\exp(-if)$, where $\tilde{\sigma}_{1,N}$ are unitary operators localized around the edges which commute with each other and with the bulk localized bits (``l-bits''); and $f$ is a local Hamiltonian acting on the bulk l-bits~\cite{sondhi_floquet_class1,sondhi_floquet_class2}. As before, the $\tilde{\sigma}_{1}\tilde{\sigma}_{N}$ piece of $U$ entails that a $\mathbb{Z}_{2}$ charge (spin flip) is pumped across our system (Fig.~\ref{fig:exp_enhancement}i) which is the entanglement spectrum signature described in the main text~\cite{vishwanath_floquet_class}.

From a more practical point of view, there is an upper bound on the disorder bandwith, $W_{\mathrm{max}}$. Instead of driving the Ising interaction term as in the ESPT case, we are toggling between two non-commuting parts, $H_1$ and $H_{2}$, of the Hamiltonian [Eqn.(3) in the main text]. Since the Floquet operator $U = \exp(-iH_{2}T/2)\exp(-iH_{1}T/2)$ then $h_{i}T/2$ should be viewed modulo $\pi/2$~\cite{khemani_phase_2015} and, therefore, $W_{\mathrm{max}}=1$. Since the amount of disorder present in the system is bounded by $W_{\mathrm{max}}$, there is an MBL-delocalization phase transition as a function of the interaction strength quantified by the power law exponent $p$ (Fig.~\ref{fig:exp_enhancement}g). Conversely, the MBL-delocalization transition can also be observed by tuning the bandwidth $W$ for a fixed power law exponent $p$, as shown in Fig.~\ref{fig:exp_enhancement}e and Fig.~\ref{fig:exp_enhancement}f.

\begin{figure}
    \centering
    \includegraphics[width=0.7\textwidth]{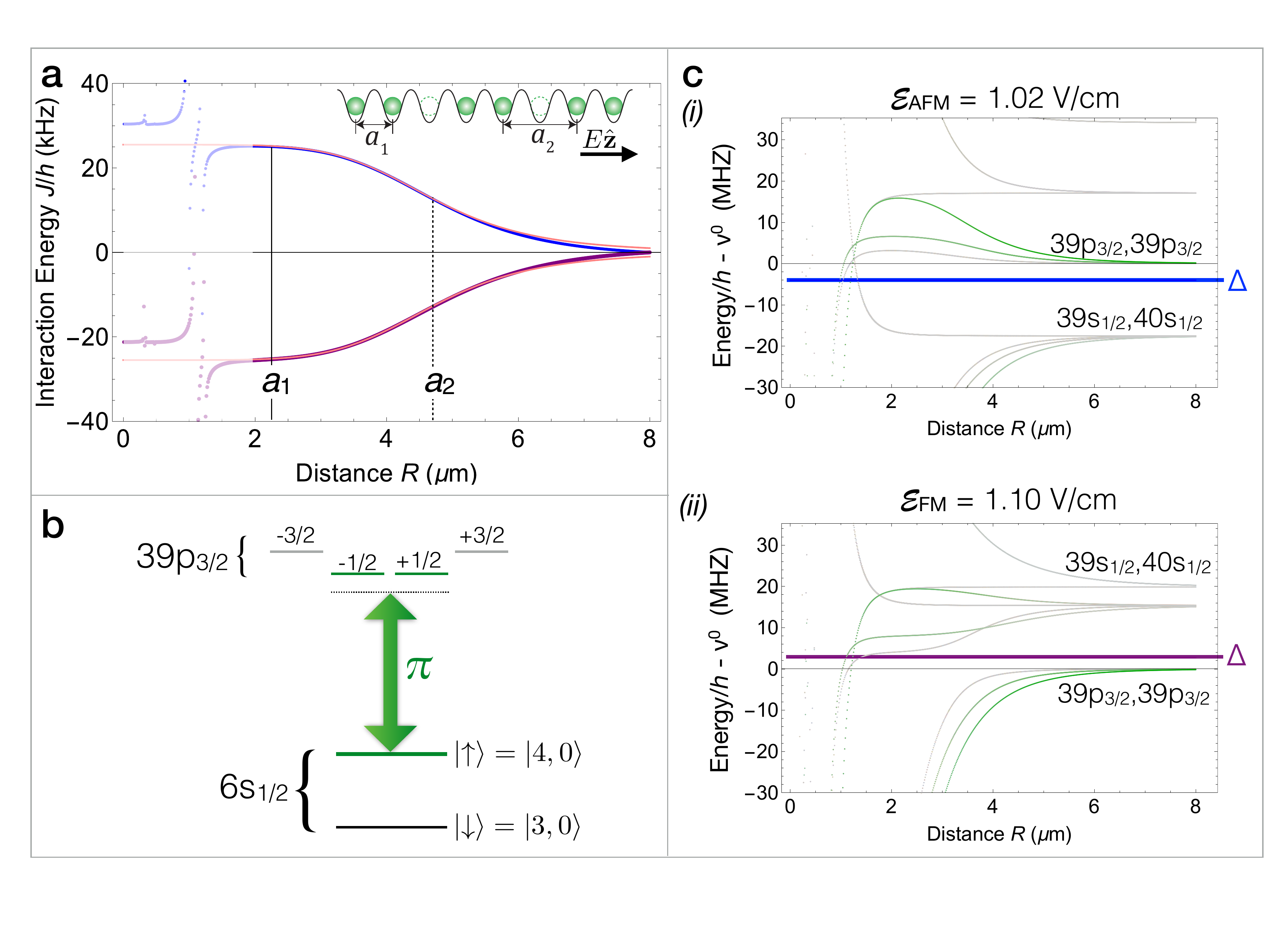}
    \caption{\textbf{Rydberg dressing with switchable sign.}  (a) Two dressed interaction potentials (blue, purple) of equal and opposite signs, for atoms in a chain along $\uvec{z}$ with interatomic spacings $\gtrsim 2~\um$.  The sign is controlled by the magnitude of an electric field in the $\uvec{z}$ direction.  The ratio of nearest to next-nearest-neighbor couplings is determined by the range of the potential relative to the lattice spacing $a_1$.  (b) Level scheme for realizing the potentials shown in (a) by single-photon Rydberg dressing in cesium.  Ising interactions are obtained by dressing only one of two ground hyperfine states.  (c) Rydberg pair potentials relevant to the dressing scheme, illustrated for two different combinations of electric field strength $\mathcal{E}$, laser detuning $\Delta$, and Rabi frequency $\Omega_{\uparrow\Ryd}$ giving rise to either (i) antiferromagnetic interactions [$\mathcal{E} = 1.02$~V/cm, $\Delta=-4$~MHz, $\Omega_{\uparrow\Ryd} = 1.19$~MHz] or (ii) ferromagnetic interactions [$\mathcal{E} = 1.10$~V/cm, $\Delta= 2.7$~MHz, $\Omega_{\uparrow\Ryd} = 0.82$~MHz].  Changing the electric field switches the sign of the F\"{o}rster defect between the $(39p_{3/2},39p_{3/2})$ and $(39s_{1/2},40s_{1/2})$ pairs, thereby changing the sign of the van der Waals interaction between $39p_{3/2}$ atoms.  Green coloring indicates the strength of coupling of each Rydberg pair state to the ground hyperfine states $\ket{\uparrow\uparrow}$ by the $\pi$-polarized laser field.  Energies are referenced to the asymptote of the Stark-shifted pair state, which differs by $\nu^0_\mathrm{AFM}-\nu^0_\mathrm{FM}=34.5$~MHz for the two different strengths of the electric field.}
    \label{fig:dressed_potentials}
\end{figure}

\section{Rydberg Dressing}

\subsection{Implementation of SPT Hamiltonians}

The proposed implementation by Rydberg dressing admits of a variety of choices for the atomic species and Rydberg state.  As an illustrative example, we here detail a dressing scheme in cesium that enables simulation of both the FSPT and ESPT phases (Fig.~\ref{fig:exp_enhancement}k and Fig.~\ref{fig:exp_enhancement}l).  While realizing the FSPT requires only a single sign of the Ising interaction (either ferromagnetic or antiferromagnetic), we present a scheme that allows for switching the sign of the interaction to enable comparison with the ESPT.

We encode the pseudo-spin in the magnetic-field-insensitive ``clock'' states in cesium: $\ket{\downarrow}=\ket{6S_{1/2}, F=3, m_F=0}$ and $\ket{\uparrow}=\ket{6S_{1/2}, F=4, m_F=0}$.  Ising interactions are introduced by a laser that couples only state $\ket{\uparrow}$ to a Rydberg state $\ket{\Ryd} = \ket{n p_{3/2}}$ at a detuning $\Delta$, as described in Refs.~\cite{Gil14,Zeiher16}.  The large hyperfine splitting ($\Delta_{\mathrm{HF}} = 2\pi\times 9.2$~GHz) between states $\ket{\uparrow}$ and $\ket{\downarrow}$ will ensure that the coupling of state $\ket{\down}$ to the Rydberg manifold is negligible for our parameters.

Toggling the sign of the Ising coupling $\pm J_{ij}$ for all interatomic distances $\abs{R_i-R_j}$ requires changing not only the sign of the detuning $\Delta$ from the Rydberg state but also the sign of the van der Waals coefficient $C_6$.  In principle, one option is to alternate between two dressing fields tuned near two different Rydberg states.  For example, in cesium, strong $C_6$ coefficients of opposite sign are obtained for Rydberg states $\ket{\Ryd} = \ket{n p_{3/2}}$ with $n = 41$ and $n=43$, proximal to a F\"{o}rster resonance between the pair states $\ket{\Ryd\Ryd}$ and $\ket{S S'}=\ket{n S_{1/2},(n+1)S_{1/2}}$ at $n=42$.  In practice, a more economical approach---requiring only a single laser field---is to dynamically control the sign of $C_6$ by using an electric field to tune the F\"{o}rster defect $E_S + E_{S'} -2E_\Ryd$~\cite{Vogt06}, where $E_\alpha$ denotes the energy of Rydberg state $\ket{\alpha}$.  The latter approach is illustrated in Fig.~\ref{fig:dressed_potentials}: a dressing laser is placed at a small detuning $\Delta$ from Rydberg state $\ket{\Ryd}=(\ket{\Ryd_+}+\ket{\Ryd_-})/\sqrt{2}$, where the degenerate states $\ket{\Ryd_\pm}\equiv \ket{39p_{3/2}, m_J=\pm 1/2}$ are shifted into F\"{o}rster resonance by an electric field oriented along the chain of atoms.  Fine-tuning the strength of the electric field switches the sign of $C_6$ and thus, in combination with changing the laser frequency, switches between antiferromagnetic and ferromagnetic dressed interactions.

The dimerization in the ESPT model is readily obtained by positioning atoms in optical tweezers with a modulated spacing $a_2 > a_1$~\cite{endres2016cold} that provides control over the coupling ratio $\lambda_2/\lambda_1$ (Fig.~1).  Alternatively, for atoms initially positioned in a lattice of uniform spacing $a_1$, ``kicking out'' every third atom such that $a_2 = 2a_1$ yields dimers with a coupling ratio $\frac{\lambda_{2}}{\lambda_{1}} = \frac{a_{1}^6 + R_{c}^6}{a_{2}^{6} + R_{c}^6}$ that can be tuned by adjusting the interaction range $R_c$ via the detuning $\Delta$ of the Rydberg dressing laser.  The ratio $\lambda_2/\lambda_1 \approx 0.45$ used in Fig.~2 is obtained by setting $0.52 R_c = a_1$.

\subsection{Calculation of Interaction Potentials}

Our method of calculating the Ising couplings generated by Rydberg dressing (Fig. \ref{fig:dressed_potentials}a) is similar to that in Ref.~\cite{vanBijnen15}.  First, we calculate $1500$ Rydberg pair potentials by diagonalizing the dipole-dipole interaction Hamiltonian $H_\mathrm{dd}$ for pair states $\ket{\alpha \alpha'}\equiv \ket{n, L, J, m_J; n', L', J', m_J'}$ with F\"{o}rster defects up to $\abs{E_\alpha + E_{\alpha'} - 2E_{\Ryd}} \lesssim h\times 70$~GHz, principle quantum numbers in the range $35\le n,n' \le 43$, and angular momentum projections $m_J + m_J' \in \{0,\pm 1\}$ accessible from the pair state $\ket{\uparrow\uparrow}$ using $\pi$-polarized light.  We account for the lowest-order effect of the electric field by including quadratic Stark shifts in the diagonal elements (F\"{o}rster defects) of $H_\mathrm{dd}$.  The resulting pair potentials for two different electric field strengths are plotted in Fig. \ref{fig:dressed_potentials}(c).  The saturation of the green color indicates the Rabi frequency
\begin{equation}
\Omega_{\psi(R)} = \sum_{\alpha,\alpha'} \langle{\psi(R)}|\alpha\alpha'\rangle\frac{\Omega_{\up\alpha}\Omega_{\up\alpha'}}{2}\left[\frac{1}{\omega_L  + (E_\uparrow - E_\alpha)/\hbar}
+\frac{1}{\omega_L + (E_\uparrow - E_{\alpha'})/\hbar}\right]
\end{equation}
of the coupling from $\ket{\uparrow\uparrow}$ to the Rydberg pair eigenstate $\ket{\psi(R)}$ at interatomic distance $R$ by absorption of two $\pi$-polarized photons of frequency $\omega_L$, in terms of the Rabi frequencies $\Omega_{\up\alpha}$ and $\Omega_{\up\alpha'}$ for single-atom excitation.

The interaction energy of the dressed pair state $\ket{\uparrow\uparrow}$ arises at fourth order in perturbation theory.  It may be understood as a reduction in the four-photon ac Stark shift
\begin{equation}
V(R) = \frac{1}{2}\sum_{\psi(R)} \frac{\abs{\Omega_{\psi(R)}}^2}{2\omega_L - E_{\psi(R)}/\hbar}
\end{equation}
when the laser becomes far off-resonant from the Rydberg pair eigenstates $\psi(R)$ of energy $E_{\psi(R)}$ due to Rydberg-Rydberg interactions.  The resulting Hamiltonian can be expressed in the form
\begin{eqnarray}
H &=& \sum_{i>j} J(\abs{R_i - R_j}) \left(\mathbb{1}+\sigma^z_i\right)\left(\mathbb{1}+\sigma^z_j\right) + \tilde{\Delta}_\mathrm{HF}\sum_{i}\sigma^z_i \\ \nonumber
&=& \sum_{i>j} J(\abs{R_i - R_j}) \sigma_{i}^{z}\sigma_{j}^{z} + \sum_{i}\left(\tilde{\Delta}_{HF} + \delta_{i} \right) \sigma_{i}^{z},
\end{eqnarray}
where $J\left(\abs{R_i - R_j}\right) = V\left(\abs{R_i - R_j}\right)-V\left(\infty\right)$; $\tilde{\Delta}_{HF}$ is the frequency of the $\ket{\downarrow}\rightarrow\ket{\uparrow}$ transition, including the ac Stark shift imparted on an isolated atom by the Rydberg dressing light; and $\delta_i = \sum_{j\neq i} J\left(\abs{R_i - R_j}\right)$ is a mean-field shift that is constant ($\delta_i \approx \delta$ for $1<i<N$) except around the edges of the chain. The inhomogeneity in $\delta_i$ can either be compensated with an additional light shift or removed via a spin-echo sequence. The Raman frequency can then simply be tuned to a constant value $(\Delta_{HF} + \delta)$.

\bibliography{SPT_biblio}